\documentclass{article}
\usepackage{ijcai22}

\usepackage{times}
\usepackage{soul}
\usepackage{url}
\usepackage[hidelinks]{hyperref}
\usepackage[utf8]{inputenc}
\usepackage[small]{caption}
\usepackage{graphicx}
\usepackage{amsmath}
\usepackage{amsthm}
\usepackage{booktabs}
\usepackage{algorithm}
\usepackage{algorithmic}
\usepackage{graphicx}
\usepackage{newtxtext,newtxmath}
\usepackage{multirow}

\usepackage{times}
\usepackage{latexsym}

\newcommand{\lqy}[1]{#1}
\usepackage{color}
\newcommand{\lqyijcai}[1]{#1}
\newcommand{\lqysql}[1]{#1}

\title{Lyra: A Benchmark for Turducken-Style Code Generation}

\author{
Qingyuan Liang$^{1,2}$\and
Zeyu Sun$^1$\and
Qihao Zhu$^{1}$\and
Wenjie Zhang$^1$\and
Lian Yu$^2$\and
Yingfei Xiong$^1$\and
Lu Zhang$^1$\footnote{Lu Zhang is the corresponding author. The Lyra dataset and code is avaliable at \url{https://github.com/LIANGQINGYUAN/Lyra}.}
\affiliations
$^1$Key Laboratory of High Confidence Software Technologies, Ministry of Education (Peking University). School of Computer Science, Peking University. Beijing, PR China\\
$^2$School of Software \& Microelectronics, Peking University. Beijing, PR China
\emails
\{liangqy, szy\_, zhuqh, zhang\_wen\_jie, lianyu, xiongyf, zhanglucs\}@pku.edu.cn
}

\begin{document}

\maketitle

\begin{abstract}
Recently, neural techniques have been used to generate source code automatically. While promising for declarative languages, these approaches achieve much poorer performance on datasets for imperative languages. Since a declarative language is typically embedded in an imperative language (i.e., the \emph{turducken}-style programming) in real-world software development, the promising results on declarative languages can hardly lead to significant reduction of manual software development efforts.

In this paper, we define a new code generation task: given a natural language comment, this task aims to generate a program in a base imperative language with an embedded declarative language. To our knowledge, this is the first turducken-style code generation task.
For this task, we present \emph{Lyra}: a dataset in Python with embedded SQL. This dataset contains 2,000 carefully annotated database manipulation programs from real-world projects. 
Each program is paired with both a Chinese comment and an English comment.
% experiment
In our experiment, we adopted Transformer, BERT-style, and GPT-style models as baselines. 
In the best setting, 
% the GPT-style model can achieve 24\% and 25.5\% AST exact matching accuracy using Chinese and English comments, respectively.
the generation performance of GPT-style models is better than others, where the AST exact matching accuracy is 24\% and 25.5\% when using Chinese and English comments, respectively.
Therefore, we believe that Lyra provides a new challenge for code generation. Yet, overcoming this challenge may significantly boost the applicability of code generation techniques for real-world software development.
\end{abstract}

\section{Introduction}

% \yxcomment{The story of the intro is a totally different one from the abstract. I suggest to use the story in the abstract.}
% 代码生成这个领域 or 问题
With the increase of software complexity, the process of programming is becoming time-consuming and error-prone ~\cite{se-1992}. 
Code generation is an important artificial intelligence problem that has the potential to release human beings from challenging software development~\cite{ATIS-1994,HS-2016}. 
It requires both understanding the meaning of development requirements and mapping them to executable source code.

% SQL, declarative, end users, high accuracy
With significant advances in deep learning, many neural-based code generation methods have been proposed. Existing approaches perform well on declarative languages including database query languages~\cite{spider-2018,wikisql-2017} and regular expressions~\cite{li2021transregex}. For example, \citeauthor{sead}~\citeyear{sead} generate SQL programs with $93\%$ execution accuracy among the test set of WikiSQL~\cite{wikisql-2017}, while \citeauthor{treegen-2020}~\citeyear{treegen-2020} achieves generate lambda calculus programs with $89\%$ exact match among the test set of ATIS~\cite{ATIS-1990,ATIS-1994}. 
These high-accuracy models facilitate end-users to perform various operations, such as querying records in a database, without understanding the syntax of the declarative language. 
Therefore, it is tempting to directly use such well performed approaches to improve the efficiency of the real-world software development. 

\begin{figure}[t]
\centering
\setlength{\abovecaptionskip}{0cm}
\includegraphics[scale=0.20]{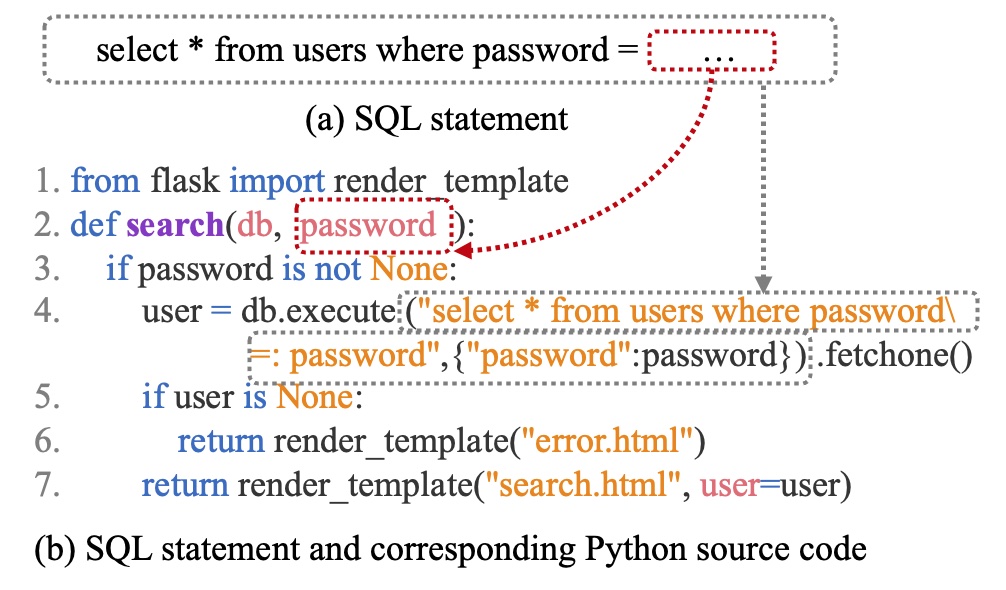}
\caption{The examples included in the Lyra dataset, (a) is the SQL statement, (b) is the source code of executing raw SQL.}
\label{motivation_example}
\end{figure}

% barely benefit for real development
% applicability
However, almost no large-scale software is written with only one declarative language. 
A declarative language is usually embedded in another imperative language, such as Java and Python. For example, as shown in Figure~\ref{motivation_example}, a SQL statement is typically embedded as a string in Python with parameters specifying the condition being queried. 
% Existing code approaches cannot generate this kind of code since it requires the knowledge of both a declarative language and a imperative language. 
Existing code generation approaches generate imperative programs with much poorer performance than generate declarative programs, making them difficult to apply to real-world software development. For example, \citeauthor{treegen-2020}~\citeyear{treegen-2020} generates Python programs with only $33\%$ human evaluated accuracy among the test set of HearthStone~\cite{HS-2016}, while CodeGPT~\cite{codexglue} generates Java programs with only $20\%$ exact match among the test set of concode~\cite{concode-2018-mapping}.

To the best of our knowledge, none of the existing datasets is suitable for the above scenario, where a declarative language is embedded in an imperative language. In fact, an existing dataset focuses on generating code in one specific language, either a declarative language or an imperative language. Therefore, it is hard to transfer the good performance on declarative languages to real-world software development. To tackle this situation, we introduce a new code generation task involving one declarative language embedded in one imperative language. Since our code generation task aims to generate a program in a base language with an embedded language for a given natural language comment, we call it \emph{turducken}-style programming. Turducken literally means a recipe method in which one animal is stuffed into the stomach of another. We use the turducken analogy to stuff one programming language into another. 
As far as we know, this is the first time to introduce the turducken-style code generation task. 

Compared with code generation for declarative languages, the turducken-style code generation is more challenging mainly for the following reasons. First, this task involves at least two different grammars. Learning two different grammars at the same time is challenging. Second, the two languages are interrelated and interdependent, and the model should have the ability to express cross-language dependency. To support this task, we construct a corresponding dataset involving Python and SQL. Python is currently a very popular programming language. Many code generation datasets are associated with Python, such as Hearthstone~\cite{HS-2016}
% , Django~\cite{Django} 
and CoNaLa~\cite{conala-2018}. There are also many datasets for text-to-SQL tasks, such as WikiSQL ~\cite{wikisql-2017} and Spider~\cite{spider-2018}. The generation of SQL on these datasets has achieved promising results~\cite{sead}.  
% ~\cite{sead,spider-top1-2021}.  

% 然而构造数据集存在挑战
Constructing a large, realistic dataset in the field of code generation is typically challenging due to the following reasons. 
First, the application scope is very wide, and the search space of programs written by different programming languages and purposes is extremely huge. 
This makes the process of collecting source code from real-world development projects far more cumbersome than collecting other data, such as images.
Second, identifying the functionality of the collected source code is also challenging.
Certain experience is necessary to understand the content and logic of the programs. 
Moreover, the code crawled from the development project often needs to be modified before it can be used. 
For example, "\texttt{\small conn = self.db.connect()}" might means to get the connection from the existing database "\texttt{\small db}". 
But "\texttt{\small self}" belongs to the dependence information outside the function block, which leads to the situation that the source code function cannot be executed independently.  
The modification of the code requires the annotator to have pre-knowledge and practical experience, and this process is essential to ensure the quality of the dataset. All of these processes are destined to be challenging to collect high-quality and realistic code generation datasets.
% 好的数据集应该具有的特点
We believe that a good code generation dataset should have satisfactory realisticness, diversity, quality and complexity. 

% The existing code generation datasets may not meet these requirements. 
% For example, the source code contained in Hearthstone~\cite{HS-2016} is all extracted from one project, resulting in limited diversity, the data of CoNaLa~\cite{conala-2018} is automatically extracted from \emph{stack overflow}, which \yxmodify{}{contains simplified code for questions and }often fails to reflect the direct needs in the development process. and Django~\cite{Django} is not undergone rigorous further processing to ensure the quality.

% 我们采取的措施
In this paper, we present a new code generation dataset, which is specific to the turducken-style code generation task. To ensure that our dataset has satisfactory characteristics, our data have been processed as follows. 
First, to meet the realisticness and diversity requirements, the source code snippets are crawled from real development projects on Github. 
Second, our data have gone through a fine annotation process of 400 human hours. Since most of the original data contain project-related information, they cannot be directly used as independent functions. 
% \yxcomment{I do not understand this example. Why must the generated code fail? Why cannot the generator learn to locate the suitable variable ``db''.} 
%To solve the challenge of data quality, we spent a lot of time to modify crawled data.
To address this challenge, we manually modified the crawled data.
% \yxcomment{The above sounds like a very adhoc change and it is not clear to me why it should be done in this way.} 
In the above case, we can put the connection information "\texttt{\small conn}" as a parameter in the parameter list of the function block.
We describe these details in our annotation process.
Additionally, to address the quality problem, we designed a quick and efficient checker to check each data. Specifically, we use some rule-based methods to check and modify the annotated data according to the characteristics of the dataset.
Finally, we constructed Lyra, which contains 2,000 carefully modified source code snippets, and each code snippet corresponds to one Chinese and one English comments.

Figure~\ref{motivation_example} shows an example in the Lyra dataset. % \yxcomment{do not use "(a)" as part of the sentence.} 
In this figure, the subfigure~(a) is a SQL statement, subfigure~(b) is the corresponding source code for executing SQL statement. 
Each source code in the dataset can be divided into three parts: preparation before SQL execution, executing SQL statements, and processing query results. 
For the example in subfigure~(b),
line 4 shows the process of executing SQL statements based on the function parameters; lines 1-3 import the package required by the function, define the function, and prepare for SQL execution; lines 5-7 process the results of the query.
Note that our data are not a simple combination of SQL statements and Python code, but there is a close interaction between them.
For example, the parameter "\texttt{\small password}" is used as the judgment condition of line 3 in the base language and also as the parameter to match the SQL string in the embedded language.
% Subfigure~(c) shows the Chinese and English comments of the source code. 
% \yxcomment{The comment seems to be too detailed to be written by a programmer who has no knowledge of the code. Yet if the programmer already knows the code, why do we need to generate it? We need a story about in what scenario the developer could produce this comment.}

\lqyijcai{In addition, we used currently popular neural network architectures, Transformer\cite{transformer} and pre-trained models~(BERT-style and GPT-style)\cite{codebert,gpt2,codexglue} both, to conduct experiments on our dataset. The best performance of the model can reach 24\% and 25.5\% AST (Abstract Syntax Tree) exact matching accuracy using Chinese and English comments, respectively.} Experimental results show that our dataset may improve the practicability of code generation tasks in some specific applications. They also suggest that our dataset is complex and there is a large room for further development and utilization.

% The paper makes the following contributions:
% \begin{itemize}
% \item  We first introduce a turducken-style code generation task. We also constructed a new dataset about using the Python programming language embedded SQL statement to manipulate the database. Each source code is provided a Chinese comment and an English comment.

% \item Our data has been elaborately annotated, and each example has gone through a rigorous quality inspection process. Each source code block is derived from a program commonly used in actual development projects.

% \item  We used the currently popular Transformer neural network architecture for experiments and used the BPE tokenization to improve the performance.  We obtain 0.5\% and 1.5\% of the code AST exact matching accuracy for Chinese and English comments. This suggests that there is a large room for further research.
% \end{itemize}

\section{Related Work}
%  Tasks similar to code generation include text-to-SQL(~\cite{wikisql-2017},~\cite{spider-2018}) and api generation tasks~\cite{deep-api-2016}, etc. 
 
 Most closely related to our task is the previous code generation task for generating programs in single programming languages~(~\cite{HS-2016,Django,sci2020,codex2021}, etc) and the text-to-SQL task for generating SQL statements~(~\cite{wikisql-2017,spider-2018,sqlsurevy2022}, etc). 
 They are all the process through the understanding of natural languages to generate the corresponding structured information. Below we discuss the research status of datasets related to these two task.

For the code generation task, several datasets with different programming languages have been created. These datasets include ATIS~\cite{ATIS-1990,ATIS-1994}, Hearthstone~\cite{HS-2016}, 
% Django~\cite{Django}, 
CONCODE~\cite{concode-2018-mapping}, and CoNaLa~\cite{conala-2018}.
% JuICe~\cite{JuICe-2019}. 
% They have been studied extensively, including approaches of LSTM~\cite{lstm-2015}, SNM ~\cite{SNM-2017}, ASN ~\cite{ASN-2017}, ReCode ~\cite{Recode-2018} and TreeGen ~\cite{treegen-2020}. 
Although deep learning-based code generation is potentially promising and existing evaluations suggest that such approaches may be more accurate~\cite{treegen-2020}, they are most evaluated on datasets where requirements are different from real-world requirements in the industry~\cite{dl-cg-review-2020}. 
On one hand, the current datasets can hardly be both of high-quality and used in the actual development process. Because the code blocks extracted from real projects often have a lot of project-related dependency information, they cannot be used directly. On the other hand, the generated programs are often significantly different from their references. They often contain syntax or semantic errors, and very little code can pass the test to meet the needs of actual use. Besides, most of the previous code generation datasets only considered one programming language, not the combination of different programming languages.

As to the text-to-SQL task, many datasets using SQL as programs have been created, such as 
% Restaurants ~\cite{Restaurants-1,Restaurants-2}, Academic ~\cite{Academic}, Scholar ~\cite{Scholar}, 
Yelp and IMDB ~\cite{Yelp_IMDB}, WikiSQL ~\cite{wikisql-2017} and Spider ~\cite{spider-2018}.
These datasets have been studied by researchers in both communities of NLP~\cite{sead} 
% ~\cite{spider-top1-2021} 
and Database ~\cite{Yelp_IMDB}. Using natural language description to generate SQL statements is easier than generating source code needed in development. 
% The evaluation method of the text-to-SQL task can also accurately and effectively judge the correctness of the generated SQL statements.
\lqysql{However, the current scenarios of text-to-SQL task are bound with end users, and in actual development, SQL statements are often embedded in specific programming languages, such as Python, to perform database manipulation.} For example, the SQL statement "\texttt{\small select id in user where name = Bob}" can only query Bob's id value. If you replace "\texttt{\small Bob}" with a variable in the base language, you can query any person specified by the variable.

That is to say, in real-world development, SQL statements are typically embedded in base languages to increase the expressiveness. 
In the field of code generation, this turducken-style task needs to considered to improve the usability of generated code.
We construct a new dataset, Lyra, about using Python to operate databases to support our task. 
% Specifically, the code in the Lyra dataset are modified source code from Github real projects using Python to operate databases. 
% Besides, the data in Lyra has undergone strict quality inspection, which way provide new ideas for the practical application of code generation in the future. 

\begin{figure}[t]
\centering
\setlength{\abovecaptionskip}{0cm}
\includegraphics[scale=0.086]{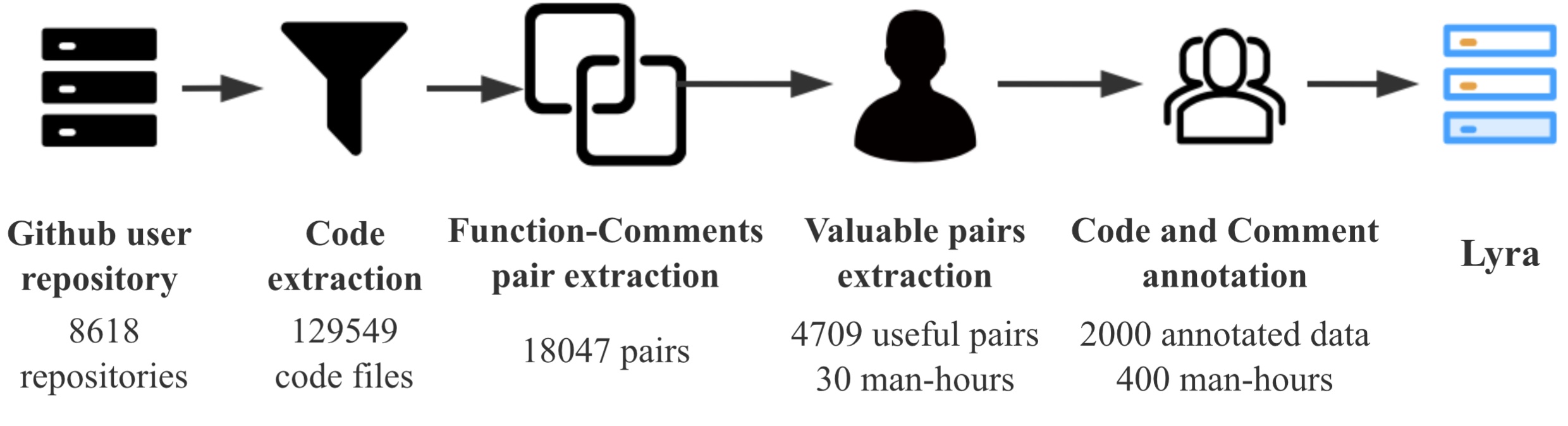}
\caption{The basic process of data set collection.}
\label{basic_process}
\end{figure}

\begin{figure}[htbp]
\centering
\setlength{\abovecaptionskip}{0.2cm}
\includegraphics[scale=0.1]{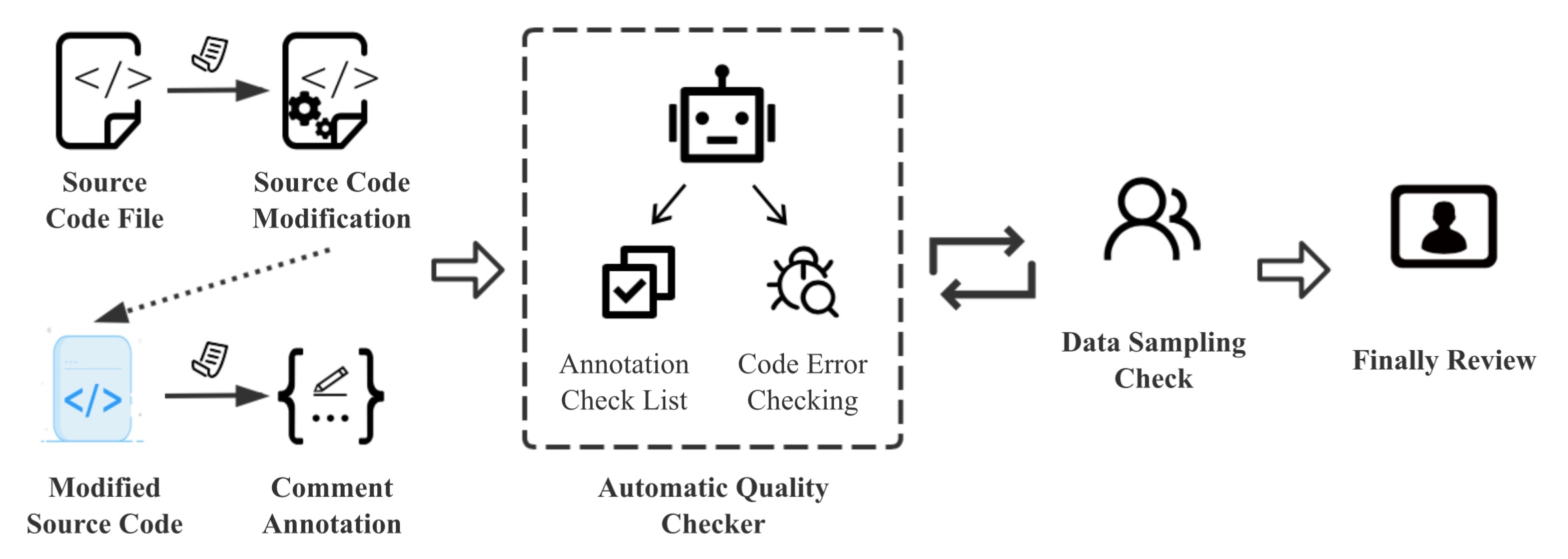}
\caption{Data annotation process, including code modification and comment annotation.}
\label{annotation_process}
\end{figure}

\section{Dataset Construction}

All the code in our dataset is crawled from Github, where each example is an independent function block. 
The source code reflects how to use Python and SQL to manipulate the database in real development. 
Ten computer science students as annotators participated in the code modification and Chinese comments annotation process, and two English professionals were responsible for annotating and checking the English comments. As illustrated in Figure \ref{basic_process}, we developed our dataset with 2,000 examples, spending around 430 hours of human labor in total. 
%ethical considerations每个人自愿加入,且对报酬满意
\lqy{Note that we first explained our annotation task and compensation plan to each candidate annotator. This ensures that candidates who intended to participate in our task knew the specific work and were satisfied with the corresponding compensation.
Finally, all participants in the annotation process were well informed and voluntary.}

\subsection{Dataset Collection}
\label{Dataset collection}
Collecting code snippets with correct functionality, clear logic, and applicable to actual development is hard. 
The code crawled from Github often has project-related operations or global variables. That results in source code not being an independent executable function block. 
Besides, the logic of some source code is often not clear enough, and even has bugs. These all bring challenges to construct the dataset.

To solve these challenges, we used the process shown in Figure \ref{basic_process} to our dataset. 
First, we obtained 8,618 user repositories from Github. Most of them are related to using the SQLAlchemy framework in Python to access the database. 
Second, we extracted 120,540 source code snippets written in the Python programming language from these repositories, and then obtained 18,047 unique function blocks related to SQL operations. 
Third, a programmer chose from these non-duplicated function blocks those that seem to have clear code logic and are easy to understand as candidates for annotation. 
Finally, we used a strict annotation process to modify the source code and annotate the corresponding comments. The final dataset contains 2,000 pieces of source code with related Chinese and English comments.

It is the most time-consuming (400 man-hours) in the code and comment annotation stage. At this stage, we first made careful modifications %according to the purpose and functionality 
of the original code so that the final code logic is clear and the functionality is correct. In addition, we used Chinese and English to annotate the code. At the end of this stage, we also checked the quality of the dataset, and conducted a sampled evaluation. The specific process at this stage is shown in Figure \ref{annotation_process}. 

\subsection{Code Modification and Comments Annotation}
\label{Annotation}
To ensure the correctness and conciseness of the code and the completeness of the comments, we have designed many rules for the annotators to follow. This stage can be divided into two parts: code modification and comments annotation. 
%ethical considerations为保证每个人被公平对待而采取的措施
\lqy{In this stage, we randomly distributed 100 pieces of data to each annotator each time. After annotation, the data need to pass the quality checking. If the code passes the quality check, we continued to provide the same amount of new data for the annotator. Everyone was treated fairly in this stage without prejudice.}
% and discrimination.
%在限制每个人最多获取数据量的情况下(每人最多500),每个人是按劳获取收益. 这样的设定使得让标注者不会由于标注数据太多而对忽略对质量的要求. 
% 因此, 在保证数据质量的同时, 也不会给标注者过多的任务压力. 
% The maximum amount of data obtained by each person is limited (500 comments or code snippets per person at most), e
\lqy{Everyone was compensated according to their workload.
Those who gave annotation with more quantity and better quality got more compensation. This setting makes the annotator not ignore the quality of annotation because of too much data.
Therefore, while ensuring the data quality, all annotators beared roughly balanced task pressure.}

% \yxcomment{This is an important part of the dataset construction, as we modify the original source. But the current description is vague, and many actions are not well-defined. For example, what is project related information? How do you determine if something is project-related and how do you remove it? How do you simplify variable names? How do you select the set of actions rather than other possible actions?}
% \paragraph{Code modification.}
\subsubsection{Code Modification}
Most of the function blocks extracted from Github are incomplete or incorrect. 
The most common problem is to refer to variables and class methods outside the function block, that is, project-related information. 
A function block containing project-related information often makes the function level code in the dataset difficult to understand.
Therefore, we must make certain modifications before the code can be used as an independent function block and can pass the static program analysis. 
Also, we need to ensure that the collected Python code is correct for using SQL to access the database. For this reason, we have added the following restrictions:
\begin{itemize}
\item Return the built-in Python data objects (list, dict,...)  after executing the SQL statement, instead of objects of other classes.
\item Keep the SQL statements whose operations are related to only the \texttt{\small SELECT} keyword, not \texttt{\small INSERT}, \texttt{\small UPDATE} and \texttt{\small CREATE}.
\item Remove project-related information and use parameters to express specific variables or classes outside the function block.
\end{itemize}

Programmers often need to generate concise code rather than an obscure code, so conciseness is very important for using the generated code in real scenarios. We set the following rules to achieve this goal: 
\begin{itemize}
\item Simplify the complex variable names (more than 30 characters). This requires the annotator to simplify according to the functionality of the source code and the meaning of the original function name.
\item Remove redundant information that does not affect the functionality of the source code, such as redundant spaces, line breaks and comments. 
\item Focus on the part of the program where Python code embedded with SQL statements. Try to remove Python code that has nothing to do with SQL operations. 

\end{itemize}

% \paragraph{Comments annotation.}
\subsubsection{Comments Annotation}
In the task of code generation, the purpose and principle of annotation are that when the programmer sees the annotation, he/she can write the code with the same functionality as the source code. 
Complete comments can generate better source code, but more complete comments is more difficult to provide. 
Ideally, we aim to generate the most useful code with the simplest description. 
In order to assist annotators to annotate the key information in the source code more effectively, we also designed some rules. 

First, for the purpose of source code annotation, we set the following principle for comment annotators: correctness, diversity, and clarity.
Correctness requires that the annotator can understand the source code well and give the corresponding code description.
Diversity and clarity are the requirements of the language level of the annotator.
At best, other programmers should be able to write the same code by referring to the annotation. 

% \yxcomment{I do not understand the motivation. What is "code composition"? Overall, I found the design rationale of the following rules unclear.}
Second, from the perspective of source code composition, we add some specific requirements.
The composition of the source code collected in this paper can be divided into temporary variables, function parameters, import functions, and built-in methods. 
We set this rule to guide the annotator to decide which content must be described and which content can be omitted.
Temporary variables in the source code do not need to be described in the comments because they may not affect the functionality of the code. The parameters in the source code need to be described in the comment and marked with the \$ symbol. Import packages and built-in methods in source code generally do not need to be described in comments if they are commonly used. But if there are rarely used, they need to be distinguished in the comments. 

Furthermore, specific to our task, we provide more detailed annotation guidance. For example, different SQL execution styles need to be distinguished in the comments, and SQL statements and their related operations should be relatively independent. These rules can make the annotator better understand the data we need for the current task.

Finally, we provided examples for each annotator to learn. In the early stage of annotation, we analyzed the examples of annotation errors and generate the corresponding documents. In addition, we used the way of real-time update and sharing to let the annotator avoid the subsequent annotation errors as much as possible.

\subsection{Automatic Quality Checker}
\label{Checker}
It is very time-consuming and error-prone to manually check the content annotated by the annotator one by one. In this paper, we designed an automated checker to check the modified code and annotated comments. 

As described in Figure~\ref{annotation_process}, our checker contains two basic components, which are content mismatch check for code comments and error detection for modified code. 
First, we designed a series of check items for the code comments. For example, we automatically check whether the comment completely describe all the parameters in the function, and check whether the comment match the corresponding source code through the characteristics of the code.
% , and use specific rules to roughly check the correspondence between Chinese and English comments.
Second, some function blocks may have specific project-related information, and cannot be used as independent function blocks. 
% Therefore, we must modify the code, and ensure that the modified code can be used as an independent function block. 
We used Pylint (https://www.pylint.org), a Python static code analysis tool, to check the code function. We check all the modified code and found some errors after modification. Finally, we returned all the problematic source code for re-modification.

\subsection{Data Sampling Check}
\label{Data Sampling Check}
We randomly selected 100 examples from the dataset each time for checking. 
We further iteratively improved the rules of the quality checker by analyzing errors from the sampled data. Until there were no obvious problems with the next random sample of data, we terminated the iterative process.

\subsection{Final Review}
\label{Review}
Finally, we asked the most experienced annotator and English professionals to conduct the final review after the data sampling check. At this stage, we tried to ensure that there are no language problems in the comments and no sensitive information in the source code. 

\section{Dataset Statistics}

We summarize the statistics of Lyra in Table \ref{dataset-statistics}. Lyra contains 2,000 source code snippets for database manipulation and their corresponding comments. There are 3 code execution styles in our dataset, which correspond to common SQL processing methods in SQLAlchemy. The first style is to execute raw SQL statements in strings. The second style is to execute SQL statements represented by Python expressions. The third style is to use SQLAlchemy's ORM (Object Relational Mapper) to execute the SQL statements. To make the style of the dataset consistent, we allow the table objects in the second and third execution-styles to be passed in as parameters. 
In addition, there is no complex SQL statement in Lyra. The SQL components involved include SELECT, COUNT, WHERE, but no complex keywords like GROUP BY, ORDER BY. 

\begin{table}
\centering
\scalebox{0.843}{
\setlength{\tabcolsep}{1mm}{
\begin{tabular}{ccllcllclc}
\toprule
\multirow{2}{*}{\textbf{Number of.}} & \multicolumn{3}{c}{\textbf{Train}}            & \multicolumn{3}{c}{\textbf{Valid}}            & \multicolumn{3}{c}{\textbf{Test}}                                 \\ \cline{2-10} 
                                     & \multicolumn{1}{l}{\textbf{mean}} & max & min & \multicolumn{1}{l}{\textbf{mean}} & max & min & \multicolumn{1}{l}{\textbf{mean}} & max & \multicolumn{1}{l}{min} \\ \midrule
Tokens of CC                         & 70.43                             & 368 & 30  & 71.18                             & 154 & 29  & 69.93                             & 157 & 30                      \\
Tokens of EC                         & 57.69                             & 202 & 25  & 58.17                             & 119 & 28  & 57.44                             & 146 & 23                      \\
AST Nodes                            & 44.36                             & 108 & 19  & 43.9                              & 129 & 23  & 43.66                             & 81  & 19                      \\
Parameters                           & 2.23                              & 6   & 1   & 2.3                               & 5   & 1   & 2.23                              & 4   & 1                       \\ \bottomrule
\end{tabular}
}
}
\caption{\label{dataset-statistics}Dataset statistics of Lyra. CC and EC represent Chinese and English comments respectively.}
\end{table}

\section{Methods}
To analyze the quality and demonstrate the purpose of our corpus, we experimented with several code generation models. 
Although many advanced methods generate code in the form of AST, they are designed to generate code from the syntax of a single programming language. \lqyijcai{These methods cannot directly be applied to our turducken-style task, so we chose some currently popular and generic neural models as baselines. 
% This section describes the details of our baselines, including Transformer, BERT-style and GPT-style models.
}

\subsection{Transformer}
\lqyijcai{Transformer~\cite{transformer} is a popular encoder-decoder framework that has surpassed RNNs on many sequence-to-sequence tasks. In the encoder, the Transformer first maps the input sequence to word embedding and position embedding. We also use word-level and BPE~\cite{bpe} methods to tokenize the source code. Then Transformer uses a stack of encoder layers for encoding. 
% Each layer has two sub-layers, namely the multi-head self-attention mechanism and the position-wise fully connected feed-forward network.
% In the decoder, we use the same method as encoding to process the raw input, and the decoder is also composed of a stack of layers. Compared with the two sub-layers in the encoder, the decoder inserts a third sub-layer that performs multi-head attention over the output of the encoder stack. 
% The final decoder output can be mapped to get the distribution of words on the vocabulary. 
Finally, the decoder outputs the word distribution on the vocabulary. 
% Note that the residual connection and layer normalization are followed in each sub-layers of the encoder and decoder.
}

\subsection{BERT-Style Models}
% codebert
% graphcodebert
\lqyijcai{
BERT-style models are pre-trained models based on the encoder in Transformer. Large pre-trained models can learn effective contextual representation from unlabeled data through self-supervised objectives and have brought significant improvement to various NLP tasks~\cite{bert}. 
% For example, BERT~\cite{bert} pre-trains Transformer's encoder with MLM(mask language modeling) and NSP(next sentence prediction) objectives and obtains new state-of-the-art results on eleven NLP tasks.
We use CodeBERT~\cite{codebert} and GraphCodeBERT~\cite{graphcodebert} as BERT-style baselines to generate code snippets. These models are pre-trained for natural language and programming language on the CodeSearchNet~\cite{codesearchnet} dataset, which contains more than 2M functions of six programming languages paired with natural language documents.
% CodeBERT is pre-trained by MLM and replaced token detection objectives on the CodeSearchNet~\cite{codesearchnet} dataset, which contains more than 2M functions of six programming languages paired with natural language documents. GraphCodeBERT introduces data flow information and is pre-trained by MLM, edge prediction, and node alignment objectives on the same dataset.
After adding the decoder structure to BERT-style pre-trained models, they can be used in generation tasks.}

\subsection{GPT-Style Models}
% GPT2
% CodeGPT
% CodeGPT-adapted
\lqyijcai{GPT-style models are pre-trained models based on the decoder in Transformer. We use GPT-2~\cite{gpt2}, CodeGPT, and CodeGPT-adapted~\cite{codexglue} as GPT-style baselines. GPT-2 is pre-trained on WebText dataset with 1.5B parameters.
% and achieves state-of-the-art results on 7 language modeling tested datasets in a zero-shot setting. 
CodeGPT shares the same model architecture with GPT-2, but CodeGPT is pre-trained from scratch on single programming language corpora in CodeSearchNet. 
% Therefore, CodeGPT has new vocabulary on code corpus and the randomly initialized model parameters.
CodeGPT-adapted is a domain-adaptive one, which take the GPT-2 model as the starting point and continually trained on code corpus. GPT-style models can perform directly on downstream tasks without adding any additional architecture. }

\section{Evaluation and Discussion}

% In this section, we first describe the evaluation metrics. Then we present the results of our experiments based on various settings.

\subsection{Evaluation Metrics}
In our experiment, we use three types of metrics to evaluate the generated code on lexical similarity, syntactic similarity, and semantic similarity, respectively. For lexical similarity, we use BLEU (bilingual evaluation understudy)~\cite{bleu} to compare the lexical similarity between the generated code and the reference code. For syntax similarity, we use Code Executable to judge whether the generated code is syntactically correct. For semantic similarity, we use AST Matching to evaluate the functionality of the generated code. Since the generated code usually contains a long SQL string and code for operating SQL, the AST Matching evaluates both the AST elements with and without SQL content, namely AST Exact Matching and AST Matching in Base Language.
% Our evaluation metrics include BLEU (bilingual evaluation understudy)~\cite{bleu}, Code Executable, AST Matching in Base Language, and AST Exact Matching. Since the generated code usually contains a long SQL string and code for operating SQL,
% our evaluation metrics both take into account the matching of the AST elements with and without SQL content. \lqy{
% % In addition, we also consider AST-related metrics and the metric about whether the generated code can be executed. 
% We use Code Executable to judge whether the generated code is syntactically correct and use the analysis of AST to evaluate the functionality of the generated code. }
% 为什么不用unit testing
\lqy{In particular, we did not use unit testing as a evaluation metric in our paper, because our code snippets are collected from Github, and it is hard to get test cases. Different from text-to-SQL datasets, our code not only comes from multiple projects but also has various parameters for each function. They cannot be executed and unit tested like SQL statements and it potentially takes heavy human involvement to construct test cases for these code snippets.}

% We release the official evaluation script along with our corpus so that the research community can share the same evaluation platform.

\paragraph{BLEU.}
\label{bleu}
The first quality metrics is BLEU. 
BLEU was initially proposed to assess the quality of machine translation~\cite{bleu}. 
% For code generation, BLEU regards the source code as an expression similar to a natural language. 
For code generation, BLEU scores are calculated to compare the lexical similarity between the generated code and the reference code, where the score is between 0 and 1. We use BLEU 4 to evaluate the generated code, which is also used to evaluate existing code generation techniques.

\paragraph{Code Executable.}
The second metric is the proportion of generated code that can be executed, in other words, can be successfully compiled. The premise of generating a functionally correct program is to ensure that the program can pass the static analysis. 
% Existing datasets rarely employ such metrics because most of the reference and generated source code cannot be compiled successfully at all, i.e., they often contain syntactic errors. 
All code in our dataset can be successfully compiled, so we use Pylint to check the generated code and calculate the ratio of the successfully complied code snippets.
% find samples without error messages during execution to calculate the ratio to the total number of test sets.

% \paragraph{AST Matching Ignoring STRING Content}
% The third quality metric is the same probability of the AST without considering the specific content of the SQL STRING in the AST. Our task requires the model to be able to generate code and a SQL string at the same time, and this metric does not consider the complete equality of the type of SQL STRING in the AST. We replaced the content of SQL STRING in the referenced and generated source code AST with a specific variable. We also replaced the other variable name in the AST, and finally calculated the equal proportion of the two ASTs.

\paragraph{AST Matching in Base Language.}
The third evaluation metric is AST matching accuracy in the base language without considering the embedded language.  
In our dataset, we calculate the AST match of the Python program without considering SQL strings. 
Specifically, we replaced the content of SQL string in the source code snippet with a specific variable before calculate the AST matching rate. In other word, we anonymized the specific variable name and SQL content. For example, 'res = conn.execute ("select id from user")' is converted to 'var\_0 = var\_1.var\_2(var\_3)'.

% \paragraph{AST Matching in Embedded Language}
% The fourth metric is AST matching accuracy in embedded language. Similar to the previous metric, we only calculate the match in embedded language, that is, the abstract syntax tree of SQL strings.

\paragraph{AST Exact Matching.}
The fifth metric is the exact match of AST, which also means that the functionality of the generated code is correct. We only replaced the variable names in the code, instead of the SQL content. The replacement of variable name does not affect the functional correctness of the function. In the above example, it is transformed into 'var\_0 = var\_1.var\_2("select id from user")'. Note that AST Exact Matching is more stringent than functional correctness, because if the code is functionally correct, the code can also be expressed in different forms.

\begin{table}
\centering
\scalebox{0.764}{
\setlength{\tabcolsep}{1mm}
\linespread{1.5}
\begin{tabular}{cccccc}
\toprule
\textbf{}      & BLEU(\%)                        & \begin{tabular}[c]{@{}c@{}}Code \\ Executable\\ (\%)\end{tabular} & \begin{tabular}[c]{@{}c@{}}AST Matching\\ in Base\\ Language(\%)\end{tabular} & \begin{tabular}[c]{@{}c@{}}AST Exact\\ Matching\\ (\%)\end{tabular} \\ 
\midrule

Transformer-EC     & 48.69    & 18.5   & 2.5   & 1   \\ 
BPE+Transformer-EC & 47.05   & 23      & 4     & 1.5     \\ 
CodeBERT-EC  & 56.72   & 51      & 8.5     & 4.5     \\ 
GraphCodeBERT-EC  & 58.61   & 46      & 12.5     & 6     \\ 
GPT-EC & \textbf{67.29}   & 88      & 24.5     & 21.5     \\
CodeGPT-EC         & 65.96   & \textbf{93}    &23.5     &21      \\ 
CodeGPT-Adapted-EC & 66.5   & 92      & \textbf{29}     & \textbf{25.5}     \\ \midrule

Transformer-CC     & 49.83    & 21    & 2      & 0   \\
BPE+Transformer-CC & 45.84   & 21.5    & 3     & 0.5     \\
GPT-CC             & 66   & 92      & 22     & 20.5 \\ 
CodeGPT-CC         & 64.88   & 91    & \textbf{26}     & \textbf{24}     \\ 
CodeGPT-Adapted-CC & \textbf{66.37}   & \textbf{96}    & 24.5     & 23     \\ \bottomrule
\end{tabular}
}
\caption{\label{performance}The performance of the Transformer. CC and EC represent Chinese and English comments respectively.}
\end{table}

% \begin{figure}[htbp]
% \centering
% \setlength{\abovecaptionskip}{0cm}
% \includegraphics[scale=0.08]{generated_example2.jpg}
% \caption{Analysis of model output using English comments on test set. (a) represents case where the generated code is AST Matching in Base Language. (b) is an example of SQL string content matching and python program not matching. (c) is the AST Exact Matching example.}
% \label{generated_example}
% \end{figure}

\subsection{Experiment Settings}
\lqyijcai{We randomly selected the 10\% of 2,000 examples in our dataset for testing and validation respectively, and the remaining 80\% for training. 
% Models are expected to generate a Python program with an embedded SQL statement, given a Chinese or English comment. 
% We report the above 4 metrics and use the AST Exact Matching as the overall evaluation metric. 
Since both CodeBERT and GraphCodeBERT use English language for pre-training, they are not used to generate code with Chinese comment. }

\subsection{Experiment Results and Discussion}
We summarize the performance of different models on our test set in Table \ref{performance}. 
% The performance of neural models includes the results on Chinese and English code comments.
\lqyijcai{Among all models, the GPT-style models perform better in our evaluation metrics. The best model in GPT-style models can reach 24\% and 25.5\% AST exact matching accuracy using Chinese and English comments.}

\lqyijcai{The generated code can be divided into three cases. In the first case, the model cannot correctly generate either SQL or Python parts in the program. In the second case, the model can only correctly generate one of the programming languages in the program, either Python or SQL. This is because the model cannot effectively learn the generation of another syntax or the model cannot learn the interaction between the two languages. For example, forgetting to generate the aggregate function of SQL or ignoring the correspondence of variables between SQL statements and Python code can lead to this situation. In the third case, the model generates code with correct functionality. This is a satisfactory situation, which means to generate the correct base program and embedded program at the same time. With the SQL in the Python program becomes complex and the processing before and after executing SQL becomes diverse, it is difficult to generate correct code.}

% In Figure \ref{generated_example}, we show several examples using English comments to generate code in the test set. Figure~\ref{generated_example}(a) shows a generated example, where the AST of Python code matches but the SQL string mismatches. Specifically, the generated code in this example ignores the aggregate function of SQL. 
% % The reason may be that there are few examples in which the aggregate function \emph{max} is expressed as \emph{largest} in the dataset. 
% Figure~\ref{generated_example}(b) shows a generated source code snippet of AST matching in the SQL string, but the Python program mismatches. Figure~\ref{generated_example}(c), BPE+Transformer generates the correct source code. These examples demonstrate that our task aims to generate the correct base program and embedded program at the same time. However, when the SQL in the Python program becomes complex and the processing before and after executing SQL becomes diverse, it is difficult to generate correct code.

% In summary, in our dataset, the quality of our data is sufficient to generate meaningful source code. But the overall performances on AST Exact Matching of all models are relatively low, indicating that our task is challenging. The experiments of more novel code generation models are worthy of further investigation on our dataset.

% 我们的baseline能够生成25的正确代码，未来可期
\lqyijcai{Although the performance of the best model in our experiment is still lower than the state-of-the-art performance on text-to-SQL tasks or text-to-Python tasks, we believe that it is promising to achieve much better performance on our dataset in the future. First, no models in our experiment exploit the characteristics or the interactions of the two programming languages. A distinctive feature of our dataset is the involvement of two different languages. While this feature may impose extra difficulty for code generation, it also provides new opportunities for approaches exploiting this feature. Second, the nature of the imperative code in our dataset is different from that in existing imperative code generation datasets. The imperative code in our dataset focuses on preparing the SQL statements and collecting query results, and does not contain complex logic. Thus, we believe that generating the imperative code alone in our dataset may technically be less difficult than generating the imperative code in existing datasets. Achieving around 25\% accuracy with a straightforward model in our experiment may have already indicated this trend.
%未来我们应该探索更多种类型的turducken风格的代码生成任务，促进代码生成技术在实际的应用
}

\section{Conclusion}
In this paper, we define a turducken-style code generation task: generating a program in a base language with an embedded language by giving a natural language comment. We also introduce Lyra, a new dataset to support our task. Lyra contains 2,000 carefully annotated database manipulation programs using the Python programming language. These data are crawled from real projects in Github and each 
source code snippet is paired with both a Chinese comment and an English comment.
Experimental results suggest that Lyra provides a new challenge for code generation. 
In future work, we plan to consider the characteristics of the two programming languages to improve the generation performance for Lyra. We also plan to explore more types of Turducken-style code generation tasks to promote the practical application of code generation, such as generating programs with JavaScript embedded in HTML and SQL embedded in XML or Java.
%不同自然语言的影响
\lqy{In addition, we also plan to explore the different effects of Chinese and English comments on generation, as well as the use of natural languages other than Chinese and English.}

\section*{Acknowledgments}
This work is sponsored by the National Key Research and Development Program of China~(No. 2017YFB1001803), and the National Science Foundation of China~(No. 61872011).

\bibliographystyle{named}
\bibliography{ijcai22}

\begin{thebibliography}{}

\bibitem[\protect\citeauthoryear{Chen \bgroup \em et al.\egroup
  }{2021}]{codex2021}
Mark Chen, Jerry Tworek, Heewoo Jun, Qiming Yuan, Henrique Ponde de~Oliveira
  Pinto, Jared Kaplan, Harri Edwards, Yuri Burda, Nicholas Joseph, Greg
  Brockman, et~al.
\newblock Evaluating large language models trained on code.
\newblock {\em arXiv preprint arXiv:2107.03374}, 2021.

\bibitem[\protect\citeauthoryear{Dahl \bgroup \em et al.\egroup
  }{1994}]{ATIS-1994}
Deborah~A. Dahl, Madeleine Bates, Michael Brown, William~M. Fisher, Kate
  Hunicke{-}Smith, David~S. Pallett, Christine Pao, Alexander~I. Rudnicky, and
  Elizabeth Shriberg.
\newblock Expanding the scope of the {ATIS} task: The {ATIS-3} corpus.
\newblock In {\em HLT}. Morgan Kaufmann, 1994.

\bibitem[\protect\citeauthoryear{Devlin \bgroup \em et al.\egroup
  }{2018}]{bert}
Jacob Devlin, Ming-Wei Chang, Kenton Lee, and Kristina Toutanova.
\newblock Bert: Pre-training of deep bidirectional transformers for language
  understanding.
\newblock {\em arXiv preprint arXiv:1810.04805}, 2018.

\bibitem[\protect\citeauthoryear{Feng \bgroup \em et al.\egroup
  }{2020}]{codebert}
Zhangyin Feng, Daya Guo, Duyu Tang, Nan Duan, Xiaocheng Feng, Ming Gong, Linjun
  Shou, Bing Qin, Ting Liu, Daxin Jiang, and Ming Zhou.
\newblock Codebert: {A} pre-trained model for programming and natural
  languages.
\newblock In Trevor Cohn, Yulan He, and Yang Liu, editors, {\em EMNLP}, pages
  1536--1547, 2020.

\bibitem[\protect\citeauthoryear{Guo \bgroup \em et al.\egroup
  }{2020}]{graphcodebert}
Daya Guo, Shuo Ren, Shuai Lu, Zhangyin Feng, Duyu Tang, Shujie Liu, Long Zhou,
  Nan Duan, Alexey Svyatkovskiy, Shengyu Fu, et~al.
\newblock Graphcodebert: Pre-training code representations with data flow.
\newblock {\em arXiv preprint arXiv:2009.08366}, 2020.

\bibitem[\protect\citeauthoryear{Husain \bgroup \em et al.\egroup
  }{2019}]{codesearchnet}
Hamel Husain, Ho-Hsiang Wu, Tiferet Gazit, Miltiadis Allamanis, and Marc
  Brockschmidt.
\newblock Codesearchnet challenge: Evaluating the state of semantic code
  search.
\newblock {\em arXiv preprint arXiv:1909.09436}, 2019.

\bibitem[\protect\citeauthoryear{Iyer \bgroup \em et al.\egroup
  }{2018}]{concode-2018-mapping}
Srinivasan Iyer, Ioannis Konstas, Alvin Cheung, and Luke Zettlemoyer.
\newblock Mapping language to code in programmatic context.
\newblock In {\em EMNLP}, 2018.

\bibitem[\protect\citeauthoryear{Li \bgroup \em et al.\egroup
  }{2021}]{li2021transregex}
Yeting Li, Shuaimin Li, Zhiwu Xu, Jialun Cao, Zixuan Chen, Yun Hu, Haiming
  Chen, and Shing-Chi Cheung.
\newblock Transregex: Multi-modal regular expression synthesis by
  generate-and-repair.
\newblock In {\em ICSE}, 2021.

\bibitem[\protect\citeauthoryear{Liang \bgroup \em et al.\egroup
  }{2022}]{sqlsurevy2022}
Qingyuan Liang, Qihao Zhu, Zeyu Sun, Lu~Zhang, Wenjie Zhang, Yingfei Xiong,
  Guangtai Liang, and Lian Yu.
\newblock A survey of deep learning based text-to-sql generation (in chinese).
\newblock {\em Sci Sin Inform}, page 1–30, 2022.

\bibitem[\protect\citeauthoryear{Ling \bgroup \em et al.\egroup
  }{2016}]{HS-2016}
Wang Ling, Phil Blunsom, Edward Grefenstette, Karl~Moritz Hermann, Tom{\'{a}}s
  Kocisk{\'{y}}, Fumin Wang, and Andrew~W. Senior.
\newblock Latent predictor networks for code generation.
\newblock In {\em ACL}, 2016.

\bibitem[\protect\citeauthoryear{{Liu} \bgroup \em et al.\egroup
  }{2020}]{dl-cg-review-2020}
H.~{Liu}, M.~{Shen}, J.~{Zhu}, N.~{Niu}, G.~{Li}, and L.~{Zhang}.
\newblock Deep learning based program generation from requirements text: Are we
  there yet?
\newblock {\em IEEE Transactions on Software Engineering}, 2020.

\bibitem[\protect\citeauthoryear{Lu \bgroup \em et al.\egroup
  }{2021}]{codexglue}
Shuai Lu, Daya Guo, Shuo Ren, Junjie Huang, Alexey Svyatkovskiy, Ambrosio
  Blanco, Colin Clement, Dawn Drain, Daxin Jiang, Duyu Tang, et~al.
\newblock Codexglue: A machine learning benchmark dataset for code
  understanding and generation.
\newblock {\em arXiv preprint arXiv:2102.04664}, 2021.

\bibitem[\protect\citeauthoryear{Oda \bgroup \em et al.\egroup }{2015}]{Django}
Yusuke Oda, Hiroyuki Fudaba, Graham Neubig, Hideaki Hata, Sakriani Sakti,
  Tomoki Toda, and Satoshi Nakamura.
\newblock Learning to generate pseudo-code from source code using statistical
  machine translation.
\newblock In {\em ASE}, pages 574--584, 2015.

\bibitem[\protect\citeauthoryear{Papineni \bgroup \em et al.\egroup
  }{2002}]{bleu}
Kishore Papineni, Salim Roukos, Todd Ward, and Wei-Jing Zhu.
\newblock Bleu: A method for automatic evaluation of machine translation.
\newblock In {\em ACL}, page 311–318, 2002.

\bibitem[\protect\citeauthoryear{Price}{1990}]{ATIS-1990}
Patti~J. Price.
\newblock Evaluation of spoken language systems: the {ATIS} domain.
\newblock In {\em HLT}, 1990.

\bibitem[\protect\citeauthoryear{Radford \bgroup \em et al.\egroup
  }{2019}]{gpt2}
Alec Radford, Jeffrey Wu, Rewon Child, David Luan, Dario Amodei, and Ilya
  Sutskever.
\newblock Language models are unsupervised multitask learners.
\newblock {\em OpenAI blog}, 1(8):9, 2019.

\bibitem[\protect\citeauthoryear{Sennrich \bgroup \em et al.\egroup
  }{2016}]{bpe}
Rico Sennrich, Barry Haddow, and Alexandra Birch.
\newblock Neural machine translation of rare words with subword units.
\newblock In {\em ACL}, 2016.

\bibitem[\protect\citeauthoryear{Sommerville}{1992}]{se-1992}
Ian Sommerville.
\newblock {\em Software engineering, 4th Edition}.
\newblock International computer science series. Addison-Wesley, 1992.

\bibitem[\protect\citeauthoryear{Sun \bgroup \em et al.\egroup
  }{2020}]{treegen-2020}
Zeyu Sun, Qihao Zhu, Yingfei Xiong, Yican Sun, Lili Mou, and Lu~Zhang.
\newblock Treegen: {A} tree-based transformer architecture for code generation.
\newblock In {\em AAAI}, pages 8984--8991, 2020.

\bibitem[\protect\citeauthoryear{Tao \bgroup \em et al.\egroup
  }{2020}]{sci2020}
Chuanqi Tao, Panpan Bao, and Zhiqiu Huang.
\newblock Code line generation based on deep context-awareness of onsite
  programming.
\newblock {\em Sci. China Inf. Sci.}, 63(9):1--3, 2020.

\bibitem[\protect\citeauthoryear{Vaswani \bgroup \em et al.\egroup
  }{2017}]{transformer}
Ashish Vaswani, Noam Shazeer, Niki Parmar, Jakob Uszkoreit, Llion Jones,
  Aidan~N. Gomez, Lukasz Kaiser, and Illia Polosukhin.
\newblock Attention is all you need.
\newblock In {\em NIPS}, pages 5998--6008, 2017.

\bibitem[\protect\citeauthoryear{Xuan \bgroup \em et al.\egroup }{2021}]{sead}
Kuan Xuan, Yongbo Wang, Yongliang Wang, Zujie Wen, and Yang Dong.
\newblock Sead: End-to-end text-to-sql generation with schema-aware denoising.
\newblock {\em arXiv preprint arXiv:2105.07911}, 2021.

\bibitem[\protect\citeauthoryear{Yaghmazadeh \bgroup \em et al.\egroup
  }{2017}]{Yelp_IMDB}
Navid Yaghmazadeh, Yuepeng Wang, Isil Dillig, and Thomas Dillig.
\newblock Sqlizer: query synthesis from natural language.
\newblock In {\em OOPSLA}, pages 1--26, 2017.

\bibitem[\protect\citeauthoryear{Yin \bgroup \em et al.\egroup
  }{2018}]{conala-2018}
Pengcheng Yin, Bowen Deng, Edgar Chen, Bogdan Vasilescu, and Graham Neubig.
\newblock Learning to mine aligned code and natural language pairs from stack
  overflow.
\newblock In {\em ICMSP}, MSR '18, page 476–486, 2018.

\bibitem[\protect\citeauthoryear{Yu \bgroup \em et al.\egroup
  }{2018}]{spider-2018}
Tao Yu, Rui Zhang, Kai Yang, Michihiro Yasunaga, Dongxu Wang, Zifan Li, James
  Ma, Irene Li, Qingning Yao, Shanelle Roman, Zilin Zhang, and Dragomir~R.
  Radev.
\newblock Spider: {A} large-scale human-labeled dataset for complex and
  cross-domain semantic parsing and text-to-sql task.
\newblock In {\em EMNLP}, pages 3911--3921, 2018.

\bibitem[\protect\citeauthoryear{Zhong \bgroup \em et al.\egroup
  }{2017}]{wikisql-2017}
Victor Zhong, Caiming Xiong, and Richard Socher.
\newblock Seq2sql: Generating structured queries from natural language using
  reinforcement learning.
\newblock {\em CoRR}, abs/1709.00103, 2017.

\end{thebibliography}

\end{document}